\documentclass[sn-mathphys]{sn-jnl}

\jyear{2022}%

\theoremstyle{thmstyleone}%
\theoremstyle{thmstyletwo}%

\theoremstyle{thmstylethree}%

\begin{document}

\title[Quantum Density Matrix]{The Quantum Density Matrix and Its Many Uses}
\subtitle{From quantum structure to quantum chaos and noisy simulators}


\author*[1]{\fnm{Apoorva D.} \sur{Patel}}\email{adpatel@iisc.ac.in}

\affil[1]{\orgdiv{Centre for High Energy Physics}, \orgname{Indian Institute of Science}, \orgaddress{\city{Bengaluru} \postcode{560012}, \state{Karnataka}, \country{India}}}


\abstract{The quantum density matrix generalises the classical concept of
probability distribution to quantum theory. It gives the complete description
of a quantum state as well as the observable quantities that can be extracted
from it. Its mathematical structure is described, with applications to
understanding quantum correlations, illustrating quantum chaos and its
unravelling, and developing software simulators for noisy quantum systems
with efficient quantum state tomography.}


\maketitle

In the textbook formulation of quantum theory, the quantum states are first
introduced as objects belonging to a Hilbert space, i.e. a complete vector
space with complex coefficients and inner product. This description is
convenient because it keeps the superposition principle of quantum
dynamics manifest, at the cost of keeping around an overall unobservable
phase. The concept of density matrix is developed later, from the outer
products of the quantum state vectors. The overall unobservable phase
disappears from it, while all the physical degrees of freedom of the quantum
state are retained. The probabilistic nature of quantum theory is easily
expressed in the density matrix formalism, and the formalism is particularly
useful in describing the behaviour of open quantum systems.

In this article, Section 1 reviews the basic structure of the quantum density
matrix \cite{nielsen,preskill}. Afterwards, several applications of the
density matrix are presented. They include the Wigner function in Section 2,
quantum chaos and its unravelling using quantum machine learning in Section 3,
and quantum simulator for an open quantum system in Section 4, with future
outlook in Section 5. The standard Dirac notation is used throughout.

\section{Basic Structure}\label{sec1}

A quantum state is a unit {\em ray} in the Hilbert space. So
$\langle\psi\vert\psi\rangle = 1$, and the whole class of vectors of the
form $e^{i\delta}\vert\psi\rangle$ is identified with the same quantum state.
The overall global phase of a quantum state is unobservable, although
relative phases between quantum states can be observed in interference
experiments. (Rays form a projective manifold consisting of equivalence
classes of vectors that differ by an overall phase, in contrast to a simpler
to work with Hilbert space, which is why quantum states are described in
the vector space language with an overall redundant phase.) Because of the
normalisation constraint and the removal of the overall phase, a quantum state
in an $n$-dimensional Hilbert space is described by $2n-2$ real parameters.

The density matrix is a quantum generalisation of the concept of the
probability distribution in statistical physics. In addition to covering
all the quantum properties that can be described in the vector space
language, it also accommodates the concept of a probabilistic ensemble.

\subsection{\bf Pure states}
Let $\{\vert i\rangle\}$ be a complete orthonormal basis in the Hilbert
space. For a quantum state $\vert\psi\rangle=\sum_i c_i\vert i\rangle$,
$\sum_i \vert c_i\vert^2=1$, also referred to as a pure state, the density
matrix is the outer product $\rho = \vert\psi\rangle\langle\psi\vert
= \sum_{ij}c_i c_j^* \vert i\rangle\langle j\vert$.
$\rho$ is Hermitian, and the normalisation condition is linear, Tr$(\rho)=1$.
Also, the overall global phase is eliminated from $\rho$ by construction,
i.e. each ray in the Hilbert space specifies a unique density matrix.
In addition, $\rho$ is also a projection operator, i.e. $\rho^2=\rho$, which
implies that its eigenvalues can be only $0$ or $1$. Taking into account
the constraint Tr$(\rho)=1$, only one eigenvalue of $\rho$ is one, while
all other eigenvalues are zero. Such a $\rho$ is also positive, because
its projection along any direction is positive,
Tr$(\rho\vert i\rangle\langle i\vert)
= \vert\langle\psi\vert i\rangle\vert^2 \ge 0$.
Altogether, a pure state density matrix for an $n$-dimensional quantum
state is described by $2n-2$ real parameters.

Upon measurement, the eigenstate $\vert i\rangle$ of the measured observable
is detected with the probability $\rho_{ii}=\vert c_i\vert^2$. Using the cyclic
property of trace, the expectation value of an observable can be expressed as
\begin{equation}
\langle O\rangle\equiv\langle\psi\vert O \vert\psi\rangle = {\rm Tr}(\rho O) ~.
\end{equation}
Furthermore, the post-measurement ensemble of quantum states can be expressed
as the transformation:
\begin{equation}
\rho \longrightarrow \sum_i P_i \rho P_i = \sum_i \vert c_i\vert^2 P_i ~,~~
P_i = \vert i\rangle\langle i\vert ~.
\end{equation}
In other words, measurement of an observable makes the density matrix
diagonal in the eigenbasis of the observable, erasing all the off-diagonal
elements that carry the superposition information. It also leads to the
ensemble interpretation of the density matrix, where multiple measurement
outcomes occur with their associated probabilities.

The Schr\"odinger evolution of the density matrix is given by:
\begin{eqnarray}
i\frac{d}{dt}\rho(t)
&=& \Big(i\frac{d}{dt}\vert\psi\rangle(t)\Big) \langle\psi(t)\vert
  + \vert\psi(t)\rangle \Big(i\frac{d}{dt}\langle\psi(t)\vert\Big) \cr
&=& H(t)\vert\psi(t)\rangle\langle\psi(t)\vert
  - \vert\psi(t)\rangle\langle\psi(t)\vert H(t) \cr
&\equiv& [H(t),\rho(t)] ~.
\end{eqnarray}
It has the formal solution: $\rho(t) = U(t,0)~\rho(0)~U^\dagger(t,0)$,
with the path-ordered evolution operator
$U(t,0) = {\cal P}(\exp(-i\int_0^t H dt))$.
This unitary evolution preserves the Hermiticity, trace and projection nature
of $\rho$.

An important property of the pure state density matrix is that it has an
inherent symplectic (i.e. phase space) structure, describable using pairs
of conjugate coordinates (see for example, Ref.\cite{mukunda}, Section 7).
This means that quantum states are never fully localised; they are smeared
objects over an area of the size of the Planck constant, for each conjugate
pair of coordinates.

\subsection{\bf Mixed states}
The properties specified by Eqs.(1-3) are linear in $\rho$, in addition to
the Hermiticity, trace and positivity conditions. Hence they hold for a
linear combination of pure state density matrices as well. Since the density
matrix is a quadratic function of $\vert\psi\rangle$, this linear combination
is not a superposition of quantum states. Rather it describes a mixture of
quantum states in a statistical ensemble. The normalisation is retained by
choosing this mixture to be a probabilistic one:
\begin{equation}
\rho_{\rm mixed} = \sum_k p_k \rho^{(k)} ~,~~ p_k \in [0,1] ~,~~
\sum_k p_k = 1 ~.
\end{equation}
The post-measurement ensemble of quantum states is such a mixture. The
probabilistic mixture nature of $\rho_{\rm mixed}$ makes it very useful
in the analysis of open quantum systems, i.e. quantum systems that are not
isolated but interact with their environments. A general $\rho_{\rm mixed}$
for an $n$-dimensional quantum state is described by $n^2-1$ real parameters.

$\rho_{\rm mixed}$ is a linear interpolation of pure state density matrices
$\rho^{(k)}$. The collection of possible $\rho_{\rm mixed}$ hence forms a
{\em convex set} with $\rho^{(k)}$ on the boundary. (In a convex set, the
complete linear interpolation between any two points of the set belongs to
the set.) $\rho_{\rm mixed}$ is positive, but it is not necessarily a
projection operator. Its eigenvalues lie in the interval $[0,1]$, so one
can write $\rho_{\rm mixed}^2 \preceq \rho_{\rm mixed}$.
In general, a particular $\rho_{\rm mixed}$ can be prepared by (infinitely)
many different combinations of $\rho^{(k)}$. Quantum theory provides no
information at all about the method of such a preparation. (This is analogous
to the situation in classical physics, where an equilibrium state can be
arrived at in many different ways, and no information about the direction
of arrival survives in the description of the equilibrium state.)

\subsection{\bf Qubit}
A qubit is a quantum state in a two-dimensional Hilbert space. Starting with
$\vert\psi\rangle = e^{i\delta}(\cos(\theta/2)\vert 0\rangle
                 + e^{i\phi}\sin(\theta/2)\vert 1\rangle)$,
$\theta\in[0,\pi]$, $\phi\in[0,2\pi]$, the density matrix for a single pure
qubit is:
\begin{eqnarray}
\rho &=& \begin{pmatrix}
         \cos^2(\theta/2) & e^{-i\phi}\sin(\theta/2)\cos(\theta/2) \cr
         e^{i\phi}\sin(\theta/2)\cos(\theta/2) & \sin^2(\theta/2)
         \end{pmatrix} \cr
     &=& \frac{1}{2} \begin{pmatrix}
         1+\cos\theta & e^{-i\phi}\sin\theta \cr
         e^{i\phi}\sin\theta & 1-\cos\theta
         \end{pmatrix}
     ~=~ \frac{1}{2} (I+\hat{n}\cdot\vec{\sigma}) ~.
\end{eqnarray}
Here $\hat{n}$ is the unit vector specifying the location $(\theta,\phi)$
on the Bloch sphere:
$\hat{n}=(\sin\theta\cos\phi,\sin\theta\sin\phi,\cos\theta)$,
and $\vec{\sigma}$ are the Pauli matrices. The geometry is illustrated in
Fig. 1. The general density matrix for a mixed state qubit lies in the
interior of the Bloch sphere, specified by the spherical polar coordinates
$(r,\theta,\phi)$. It has the form
$\rho_{\rm mixed} = \frac{1}{2} (I+\vec{r}\cdot\vec{\sigma})$ with $r\in[0,1]$.

\begin{figure}[t]
\begin{center}
{\setlength{\unitlength}{1cm}
\begin{picture}(8,7)
\thicklines
\put(4,4){\vector(1,0){3}}
\put(4,4){\vector(0,1){3}}
\put(4,4){\vector(-2,-1){2.5}}
\multiput(4,4)(0,-0.5){6}{\line(0,-1){0.3}}
\put(1.5,3){\makebox(0,0)[bl]{$x$}}
\put(6.8,4.2){\makebox(0,0)[bl]{$y$}}
\put(4.2,6.8){\makebox(0,0)[bl]{$z$}}

\put(4,6){\circle*{0.15}}
\put(3.5,6.1){\makebox(0,0)[bl]{$\vert 0\rangle$}}
\put(4,2){\circle*{0.15}}
\put(3.5,1.5){\makebox(0,0)[bl]{$\vert 1\rangle$}}

\qbezier(2,4)(2,6)(4,6)
\qbezier(4,6)(6,6)(6,4)
\qbezier(6,4)(6,2)(4,2)
\qbezier(4,2)(2,2)(2,4)

\put(4,4){\vector(1,1){1.1}}
\put(5.1,5.1){\circle*{0.15}}
\put(5.2,5){\makebox(0,0)[bl]{$\vert\psi\rangle$}}
\put(4,4){\vector(3,1){1.7}}
\put(5.5,4.2){\makebox(0,0)[bl]{$\hat{m}$}}

\thinlines
\qbezier(4,6)(5.2,6)(5.3,4)
\qbezier(5.3,4)(5.2,2)(4,2)
\multiput(4,4)(0.8,-0.4){2}{\line(2,-1){0.4}}

\qbezier(4,4.4)(4.2,4.5)(4.3,4.3)
\put(4.15,4.5){\makebox(0,0)[bl]{$\theta$}}
\qbezier(3.35,3.65)(4,3.4)(4.65,3.65)
\put(4.1,3.2){\makebox(0,0)[bl]{$\phi$}}

\put(4,4){\vector(-1,-1){1.1}}
\put(2.9,2.9){\circle*{0.15}}
\put(3.1,2.7){\makebox(0,0)[bl]{$\vert\psi_\perp\rangle$}}

\end{picture}
}
\end{center}
\vspace{-1cm}
\caption{The Bloch sphere representation of a qubit. The basis states
$\vert 0\rangle$ and $\vert 1\rangle$ correspond to the north and the
south poles respectively. A rotation by angle $\theta$ around the direction
$\hat{m}$ takes the state $\vert 0\rangle$ to the state $\vert\psi\rangle$.
The state $\vert\psi_\perp\rangle$ orthogonal to $\vert\psi\rangle$
corresponds to its diametrically opposite point.}
\end{figure}
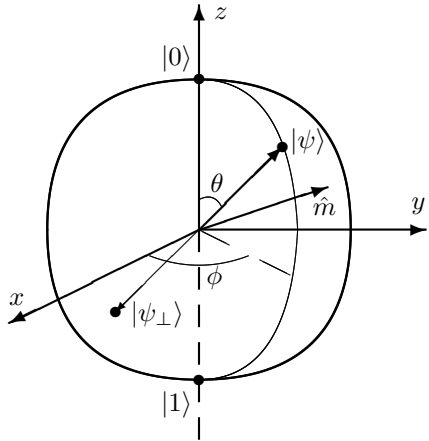

\subsection{Reduced density matrix}
Bipartite quantum systems provide a setting to illustrate many quantum
puzzles, arising from the unusual properties of quantum correlations.
The setting is versatile enough to tackle a variety of situations:
two qubits, two quantum registers, a quantum system and its measurement
apparatus, and a quantum system and its environment (i.e. the rest of the
universe). It must be kept in mind that the correlations depend on how
the whole system is divided into two parts.

Many questions about bipartite systems concern what can be learned from
transforming or observing only one part. In a probabilistic framework,
the calculations for such instances can be simplified by summing over all
possibilities for the part that remains unobserved. Let the whole Hilbert
space be ${\cal H}_A\otimes{\cal H}_B$, transformations that act only on
part $B$ have the form $I_A\otimes U_B$, and operators that measure
properties of only part $B$ have the structure $I_A\otimes O_B$. On the
other hand, the density matrix describing the state of the whole system,
$\rho_{AB}$, may not factorise due to correlations.

The generic density matrix for the whole system can be expanded in terms
of orthonormal bases for parts $A$ and $B$ as:
$\rho_{AB} = \sum_{ijkl} \rho_{ik,jl}
\vert i_A\rangle\vert j_B\rangle\langle k_A\vert\langle l_B\vert$.
Then the reduced density matrix for the part $B$, corresponding to a sum
over all the unobserved possibilities of the part $A$, is the partial
trace of $\rho_{AB}$ over part $A$:
\begin{equation}
\rho_B = {\rm Tr}_A(\rho_{AB})
       = \sum_i \rho_{ii,jl} \vert j_B\rangle\langle l_B \vert ~.
\end{equation}
By construction, $\rho_B$ is Hermitian, positive and ${\rm Tr}_B(\rho_B)=1$.
Under a local transformation of part $B$, it evolves to $U_B\rho_B U_B^\dagger$.
The expectation value for observing a local operator on part $B$ is:
\begin{equation}
\langle O_B\rangle = {\rm Tr}_{AB}(\rho_{AB}(I_A\otimes O_B))
= {\rm Tr}_B(\rho_B O_B) ~.
\end{equation}
$\rho_B$ is indeed a mixed state density matrix, and it is not always a
projection operator. This is a generic result---the change in the
nature of the density matrix is not a dynamical process, but it is a
consequence of ignoring some degrees of freedom of the whole system.

With a specific basis choice, superposition of the basis states produces
off-diagonal elements for the density matrix. These off-diagonal elements are
complex numbers in general, and can interfere destructively when summed over
different possibilities. The off-diagonal elements are called coherences,
and their suppression is known as {\em decoherence}. When a quantum system
interacts with its environment, it is impossible to keep track of the
environmental degrees of freedom and hence they are summed over, which
suppresses the coherences and drives the density matrix towards a diagonal
form. A diagonal density matrix corresponds to a classical probability
distribution, and decoherence provides a means to understand how a quantum
system can reduce to a classical one. It should be kept in mind that the
nature of the system-environment interaction determines what would be the
appropriate diagonal basis, which is also referred to as the preferred basis.

\subsection{Gleason's theorem}
The axiomatic formulation of quantum mechanics assumes a description of
states and operators, and then provides a prescription for probabilistic
outcomes of operator measurements. Gleanson's theorem provides a powerful
counter-statement that any theory obeying certain rules of probabilistic
observations (which may be quantum or classical) for all its states must
have a description in terms of a density matrix with specific properties.
Together, they make the density matrix formalism of quantum mechanics
both necessary and sufficient.

Consider the situation where independent measurement settings $\{M_i\}$
for a system produce outcomes with probabilities $\{p_\phi(M_i)\}$ for
the state $\phi$. Then it is logical to assume that:\\
(i) Probability that no outcome is measured is zero, i.e. $p_\phi(0)=0$.\\
(ii) Probability of all possible outcomes is one, i.e.
$p_\phi(\sum_i M_i)=1$.\\
(iii) Probabilities of independent outcomes add, i.e.
$p_\phi(M_i+M_j)=p_\phi(M_i)+p_\phi(M_j)$ for $i\ne j$.\\
In the Hilbert space, any measured operator can be expressed
in terms of a complete orthonormal set of projection operators,
$O=\sum_i \lambda_i P_i$ with $\sum_i P_i=I$. Then the orthogonality of
the projection operators $\{P_i\}$, i.e. $P_iP_j=0$ for $i\ne j$, allows
them to be identified as the measurement settings $\{M_i\}$.

In these circumstances, Gleason's theorem guarantees that there exists a
unique solution in terms of a Hermitian, positive and unit-trace density
matrix such that $p_\phi(P_i) = {\rm Tr}(\rho(\phi) P_i)$, for any Hilbert
space of dimension greater than two. The key assumption here is that
probabilities add for independent measurement outcomes. With the freedom
to choose the basis directions in a Hilbert space, the theorem can be
interpreted as applying either to all states for a fixed $\{P_i\}$, or
all possible $\{P_i\}$ for a fixed state. Note that the classical
situation just corresponds to a diagonal density matrix.

An exception occurs for the two-dimensional Hilbert space of a qubit,
because it does not have enough orthogonal directions and the assumption
(iii) reduces to $p_\phi(M_1)+p_\phi(M_2)=1$. The density matrix solution
remains valid, but it is not unique and (infinitely) many other solutions
can also be constructed.

These properties let us look upon the density matrix as a quantum
generalisation of the classical probability distribution. The off-diagonal
elements, which can be complex, bring in new features for the behaviour of
expectation values that are absent in the classical version.
(It is worth remembering that complex numbers were invented to solve quadratic
equations that did not have real solutions, and turned out to be powerful
enough to obtain roots of polynomials of any order.)

\subsection{Schmidt decomposition}
This is a striking result of linear algebra, which predates quantum theory.
It simplifies the description of correlations between two complementary parts
of a quantum system, by making a clever choice of basis.

Any pure quantum state of a bipartite system can be expressed in the form:
\begin{equation}
\vert\psi_{AB}\rangle
= \sum_{i,\mu} a_{i\mu}\vert i_A\rangle\vert\mu_B\rangle
\equiv \sum_i \vert i_A\rangle\vert\overline{i}_B\rangle ~,
\end{equation}
where $\vert i_A\rangle\in{\cal H}_A$ and $\vert\mu_B\rangle\in{\cal H}_B$
form complete orthonormal bases, while the vectors
$\vert\overline{i}_B\rangle \equiv \sum_\mu a_{i\mu}\vert\mu_B\rangle \in {\cal H}_B$
may not be either normalised or mutually orthogonal. Now choose the
orthonormal basis $\{\vert i_A\rangle\}$ such that the reduced density
matrix $\rho_A$ is diagonal. $\rho_A$ can be also expressed as the partial
trace ${\rm Tr}_B(\rho_{AB})$. Comparison of the two forms gives:
\begin{eqnarray}
\rho_A &=& \sum_i p_i\vert i_A\rangle\langle i_A\vert \cr
&=& {\rm Tr}_B\Big( \big(\sum_i\vert i_A\rangle\vert\overline{i}_B\rangle\big)
    \big(\sum_j\langle j_A\vert\langle\overline{j}_B\vert\big) \Big)
= \sum_{ij} \langle\overline{j}_B\vert\overline{i}_B\rangle~\vert i_A\rangle\langle j_A\vert ~.
\end{eqnarray}
Consistency in the orthonormal basis $\{\vert i_A\rangle\}$ requires that
$\sum_j \langle\overline{j}_B\vert \overline{i}_B\rangle = p_i\delta_{ij}$.
Thus $\{\vert i_B\rangle\}$ also form an orthogonal basis, and the vectors
$\vert i_B'\rangle = p_i^{-1/2}\vert\overline{i}_B\rangle$ are orthonormal.
Moreover, we can also express
$\vert\psi_{AB}\rangle = \sum_i p_i^{1/2} \vert i_A\rangle\vert i_B'\rangle$,
and have the reduced density matrix
$\rho_B = \sum_i p_i\vert i_B'\rangle\langle i_B'\vert$.

This result, which converts a bipartite quantum state from a double sum over
indices $i$ and $\mu$ to a single sum over index $i$, by a clever choice of
basis, has many physical implications (subject to the specific choice of
partition):\\
$\bullet$
There is no restrictions on the dimensionalities of ${\cal H}_A$ and 
${\cal H}_B$. The number of non-zero values of $p_i$ that appear in the
preceding expansions of the reduced density matrices $\rho_A$ and $\rho_B$
is called the Schmidt rank $r_S$. Obviously,
$r_S \le {\rm min}({\rm dim}({\cal H}_A),{\rm dim}({\cal H_B}))$.
When $r_S=1$, the quantum state factorises between parts $A$ and $B$,
and there are no correlations. But when $r_S>1$, the quantum state does
not factorise between parts and $A$ and $B$, and such states are called
{\em entangled}.\\
$\bullet$
When ${\rm dim}({\cal H}_A)\le {\rm dim}({\cal H}_B)$, only up to
${\rm dim}({\cal H}_A)$ degrees of freedom of ${\cal H}_B$ can be correlated
with those of ${\cal H}_A$. This is true even if ${\cal H}_B$ has many more
degrees of freedom than ${\cal H}_A$, as is often the case when $A$ labels
the system and $B$ its environment. Diagonalisation of $\rho_B$ is needed to
explicitly find these degrees of freedom, but diagonalisation of $\rho_A$ is
enough to specify their number. The correlations are constrained by the
one-to-one correspondence between $\vert i_A\rangle$ and $\vert i_B'\rangle$,
and that is known as {\em monogamy}.\\
$\bullet$
The orthonormal basis sets $\{\vert i_A\rangle\}$ and $\{\vert i_B'\rangle\}$
with non-zero values of $p_i$ have the same size. So they can be related by
a unitary transformation (including both rotations and reflections). Also,
the Schmidt decomposition is unaffected by independent local unitary
transformations on the two parts. Any transformation of the form
$U_A\otimes U_B$ merely redefines the basis sets $\{\vert i_A\rangle\}$
and $\{\vert i_B'\rangle\}$.\\
$\bullet$
Since any mixed state density matrix can be diagonalised as
$\rho_A = \sum_i p_i\vert i_A\rangle\langle i_A\vert$, it can always be
extended to a pure state by adding suitable $\vert i_B'\rangle$. Such an
extension of a mixed state to a pure state is not unique, but the required
number of $\vert i_B'\rangle$ does not exceed dim$({\cal H}_A)$, and so the
pure state dimension does not exceed $({\rm dim}({\cal H}_A))^2$. This concept
turns out to be very useful in construction of error-correction codes for
bounded error quantum computation that eliminate undesired system-environment
correlations. It is also useful in construction of error mitigation schemes
that focus on removing the dominant errors corresponding to non-leading large
$p_i$.\\
$\bullet$
We now have the framework to start with a mixed quantum state, extend it to
a pure quantum state, evolve it through unitary transformations, and then
perform projective measurements as well as sum over unobserved degrees of
freedom to get back a mixed quantum state. These steps can be merged to
construct a direct quantum evolution of the initial mixed state to the
final one. Such a merged description is called a {\em superoperator} or a
{\em quantum channel} or a {\em completely positive trace preserving} (CPTP)
map. In it, the quantum state is not a ray, its evolution is not unitary,
and its measurements are not projective. Such a generalised framework is
useful in the study of open quantum systems and decoherence. The
infinitesimal time evolution form of the CPTP map, with some additional
assumptions, gives {\em master equations}.\\
$\bullet$
The correlations between the two parts of a pure quantum state can be
quantified in terms of the entropy:
\begin{equation}
S(\{p_i\}) = -\sum_i p_i \log(p_i)
= -{\rm Tr}(\rho_A\log(\rho_A)) = -{\rm Tr}(\rho_B\log(\rho_B)) ~.
\end{equation}
Noting that $S(\vert\psi_{AB}\rangle) = -{\rm Tr}(\rho_{AB}\log(\rho_{AB}))
= 0$ for the pure state $\vert\psi_{AB}\rangle$, $S(\{p_i\})$ is called the
{\em entropy of formation} of the mixed state. $S(\{p_i\})$ is maximised
when all $p_i$ are equal, $S_{\rm max}=\log(r_S)$. That corresponds to
equipartition or the microcanonical ensemble of statistical mechanics.\\
$\bullet$
For a system of two qubits, the Schmidt decompostion is
$\vert\psi_{AB}\rangle = \sqrt{p} \vert i_A\rangle\vert i_B'\rangle
                       + \sqrt{1-p} \vert j_A\rangle\vert j_B'\rangle$,
with $p\in[0,1]$ and $i\ne j$. In this case, the entropy $S(p)$ is a
monotonically increasing function of $p$ for $p\in[0,\frac{1}{2}]$, and
can be used to compare correlations between the two qubits, i.e. specify
whether one two-qubit system is more or less correlated than another one.
The choice $p=\frac{1}{2}$ gives the maximally entangled Bell states,
which form a complete orthonormal basis in the four-dimensional Hilbert
space. With the one-to-one correspondence between $\vert i_A\rangle$
and $\vert i_B'\rangle$, they are very useful in construction of quantum
cryptographic protocols.

\subsection{Superoperator evolution}
Consider the generic evolution of a fully specified quantum state $\rho_A$,
which is initially uncoupled from its environment. Since $\rho_A$ is
a linear combination of pure states in the ensemble interpretation, it
suffices to consider action of a generic $U_{AB}$ on the quantum state
$\vert\psi_A\rangle\otimes\vert 0_B\rangle$. Then, in terms of a complete
set of basis $\{\vert\mu_B\rangle\}$,
\begin{equation}
U_{AB}(\vert\psi_A\rangle\otimes\vert 0_B\rangle)
= \sum_\mu \vert\mu_B\rangle\langle\mu_B\vert U_{AB}
  \vert\psi_A\rangle\otimes\vert 0_B\rangle
= \sum_\mu M_\mu \vert\psi_A\rangle\otimes\vert\mu_B\rangle ~,
\end{equation}
with $M_\mu = \langle\mu_B\vert U_{AB}\vert 0_B\rangle$.
Unitarity of $U_{AB}$ for any $\vert\psi_A\rangle$ implies that
$\sum_\mu M_\mu^\dagger M_\mu = I$. The corresponding density matrix
evolution is:
\begin{equation}
\rho_A'= {\rm Tr}_B(U_{AB}~\rho_{AB}~U^\dagger_{AB})
= \sum_\mu M_\mu~\rho_A~M^\dagger_\mu ~,
\end{equation}
which maintains Hermiticity, trace and positivity.

This operator-sum representation (or Kraus representation) is extremely
useful in analysis of open quantum systems. In particular:\\
$\bullet$
Both unitary evolution and projective measurement are its special cases.
The former has only one term in the sum, while the latter replaces $M_\mu$
by the projection operators $P_\mu$.\\
$\bullet$
Generalised measurement is the reduction to ${\cal H}_A$ of a projective
measurement in ${\cal H}_A\otimes{\cal H}_B$. It defines a positive operator
valued measure (POVM) with probabilities $p_a = {\rm Tr}(\rho\Pi_a)$,
$\sum_a \Pi_a = I$. The operators $\Pi_a$ need not be normalised or
orthogonal, but they can be expressed as
$\Pi_a = \lambda_a\vert a\rangle\langle a\vert$, $\lambda_a\ge0$.
The resultant density matrix evolution is also a particular case of Eq.(12).
\begin{equation}
\rho\longrightarrow\rho' = \sum_a\sqrt{\Pi_a}~\rho~\sqrt{\Pi_a} ~.
\end{equation}
$\bullet$
The operator-sum representation is not unique. The Kraus operators $M_\mu$
can be traded for $N_\mu=\sum_\nu U_{\mu\nu}M_\nu$ by a unitary change of
basis. The number of independent Kraus operators, however, cannot exceed
$(({\rm dim}({\cal H}_A))^2-1)(({\rm dim}({\cal H}_B))^2-1)$, in terms of 
the number of elements of $\rho_A$ and $\rho_B$.\\
$\bullet$
The superoperator evolution is reversible only when it is unitary. Otherwise,
it defines a semigroup, with the decoherence providing an arrow of time.
When $\sum_\mu M_\mu M_\mu^\dagger = I$ as well, the quantum channel is
called {\em unital}, and the entropy of the system increases monotonically.\\
$\bullet$
The linearity of the superoperator evolution can be justified by the
ensemble interpretation. But a stronger property than positivity, called
complete positivity, is required to make it a legitimate description in
the whole universe. This property states that any extension of ${\cal H}_A$
to ${\cal H}_A\otimes{\cal H}_B$ must be positive (not just a specific
${\cal H}_B$ represented by $\vert 0_B\rangle$ before). An example of a map
that is positive but not completely positive is $\rho\rightarrow\rho^T$.\\
$\bullet$
The Kraus representation theorem shows that any superoperator evolution
preserving linearity, Hermiticity, trace and complete positivity, has the
form given by Eq.(12), and the only freedom is a unitary change of basis
for $M_\mu$. Importantly, for the purpose of describing the evolution of
the density matrix, there is no loss of generality in assuming that the
initial quantum state is uncoupled from its environment (even when it may
not be true in reality).\\
$\bullet$
The operator-sum representation can be converted to a differential local
time evolution with the Markovian approximation. This approximation amounts
to assuming that the information that leaks to the environment does not
return to the system during the time scales considered, or equivalently
the equilibration time scale of the environment is much shorter than the
evolution time scale of the system being considered. Then considering
evolution for time $dt$, one can write
\begin{equation}
M_0 = I + (-iH+K) dt ~,~~ M_{\mu\ne0} = L_\mu\sqrt{dt} ~.
\end{equation}
The Hamiltonian in $M_0$ is chosen to agree with Eq.(3), and the Kraus operator
completeness relation fixes $K=-\frac{1}{2}\sum_{\mu>0} L_\mu^\dagger L_\mu$.
The resultant evolution is the Gorini-Kossakowski-Sudarshan-Lindblad master
equation:
\begin{equation}
i\frac{d}{dt}\rho(t) = [H(t),\rho(t)] + i\sum_{\mu>0}
\Big( L_\mu\rho L_\mu^\dagger - \frac{1}{2}L_\mu^\dagger L_\mu\rho
                              - \frac{1}{2}\rho L_\mu^\dagger L_\mu \Big) ~.
\end{equation}

\section{Wigner Function}

Wigner function is just the density matrix in the representation, where
one relative index is Fourier transformed to its conjugate variable
\cite{wigner}. It therefore encodes complete information of a quantum
system. It is real by construction. Since it is defined in the symplectic
phase space, its domain is quantised in units of the Planck constant.

\subsection{Infinite dimensional systems}

The Wigner function for a continuous one-dimensional quantum state is:
\begin{eqnarray}
W(x,p) &=& {1\over 2\pi\hbar} \int_{-\infty}^{\infty} dy ~
           \psi^*(x-{y\over2}) e^{ipy/\hbar} \psi(x+{y\over2}) \nonumber\\
       &=& {1\over 2\pi\hbar} \int_{-\infty}^{\infty} dy ~
           \rho(x-{y\over2},x+{y\over2}) e^{ipy/\hbar} ~,
\end{eqnarray}
\begin{equation}
\rho(x-{y\over2},x+{y\over2}) = \int_{-\infty}^{\infty}
                                dp~W(x,p) e^{-ipy/\hbar} ~.
\end{equation}
It can be negative, but its marginals are non-negative.
\begin{equation}
\int_{-\infty}^{\infty} dp~W(x,p) = \vert\psi(x)\vert^2 = \rho(x,x) ~,~~
\int_{-\infty}^{\infty} dx~W(x,p) = \vert\tilde{\psi}(p)\vert^2 ~.
\end{equation}
Its smeared values over a phase space volume element
$\Delta x \Delta p=2\pi\hbar$ (associated with counting
of states in quantum statistics) are also non-negative.
The normalisation condition is:
\begin{equation}
{\rm Tr}(\rho)=1 \quad\longleftrightarrow\quad
\int_{-\infty}^{\infty} dx~dp~W(x,p) = 1 ~.
\end{equation}

The expectation value of a Hermitian operator $O$ is obtained as:
\begin{eqnarray}
\langle O\rangle \equiv {\rm Tr}(\rho O)
&=& \int dx~dy~\rho(x-{y\over2},x+{y\over2})~O(x+{y\over2},x-{y\over2}) \nonumber\\
&=& \int dx~dy \int dp~W(x,p)e^{-ipy/\hbar} \int dq~O(x,q)e^{iqy/\hbar} \nonumber\\
&=& 2\pi\hbar \int dx \int dp~W(x,p) \int dq~O(x,q)~\delta(p-q) \nonumber\\
&=& 2\pi\hbar \int dx~dp~W(x,p)~O(x,p) ~.
\end{eqnarray}
It should be noted that $O(x,p)$ implicitly defined here is Hermitian,
and its normalisation is fixed by the convention $\langle I\rangle=1$.

\subsection{Finite dimensional systems}

For a finite dimensional quantum system with $d$ degrees of freedom,
the odd and even values of $d$ need to be handled separately.
When $d$ is odd,
\begin{equation}
W(n,k) = {1\over d} \sum_{m=0}^{d-1} \rho_{n-m,n+m} e^{4\pi ikm/d} ~,
\end{equation}
is a valid Wigner function \cite{dimodd,primefac}. Here the indices are
defined modulo $d$, i.e. $n,k,m\in Z_d = \{0,1,...,d-1\}$. The odd value
of $d$ allows all independent indices to be covered in two cycles of $Z_d$.

This definition does not work for even $d$. If the index shift is made
one sided, the Wigner function does not remain real. So an alternative
construction is needed, incorporating a ``quantum square-root".
Since any integer is an odd number times a power of two, figuring out
the Wigner function for $d=2$ (i.e. a qubit) is sufficient to reach any
$d$ using tensor products.

For $d=2$, the Wigner function can be defined using eigenvalues of
$\sigma_z$ and $\sigma_x$ as the two conjugate labels (replacing $x$
and $p$). $\sigma_z$ and $\sigma_x$ are related by the Hadamard operator,
$\sigma_z = H\sigma_xH$, which gives the discrete Fourier transformation
in $d=2$. For instance, one can call $W(+,+)$ the weight for the spin
being up along both $z$-axis and $x$-axis. The Wigner function for a
qubit can be constructed as a map from the Bloch sphere representation,
$\rho=(I+\hat{n}\cdot\vec{\sigma})/2$, with the replacements:
\begin{eqnarray}
I ~\rightarrow~ {1\over2}\begin{pmatrix} 1 & 1\cr 1 & 1 \end{pmatrix} &,&~~
\sigma_x ~\rightarrow~ {1\over2}\begin{pmatrix} 1 & -1\cr 1 & -1 \end{pmatrix} ~,\cr
\sigma_y ~\rightarrow~ \pm{1\over2}\begin{pmatrix} 1 & -1\cr -1 & 1 \end{pmatrix} &,&~~
\sigma_z ~\rightarrow~ {1\over2}\begin{pmatrix} 1 & 1\cr -1 & -1 \end{pmatrix} ~.
\end{eqnarray}
The ambiguity in the sign for $\sigma_y$ is related to the
charge conjugation symmetry of the $SU(2)$ group algebra,
$\vec{\sigma}\leftrightarrow-\vec{\sigma^*}$,
and both choices should be checked for consistency.

The normalisation condition Tr$(\rho)=1$ becomes $\sum_{ij}W(i,j)=1$,
while $\rho^2\preceq\rho$ gives $\sum_{ij}W(i,j)^2 \le 1/2$. A simple set of
qubit Wigner function weights, for a pure state allowing negative values,
is $(0.6,0.3,0.2,-0.1)$ \cite{feyn_neg}.

The expectation values can be expressed as
$\langle O\rangle = \sum_{ij}W(i,j)~O(i,j)$,
where the operator normalisation, fixed by imposing $\langle I\rangle=1$,
is different from that for the Wigner function. The qubit operators map as:
\begin{equation}
I ~\rightarrow~ \begin{pmatrix} 1 & 1 \cr 1 & 1 \end{pmatrix} ~,~
\vec{\sigma}\cdot\hat{n} ~\rightarrow~
\begin{pmatrix} n_x \pm n_y + n_z ~&~ -n_x \mp n_y + n_z \cr
                n_x \mp n_y - n_z ~&~ -n_x \pm n_y - n_z \end{pmatrix} ~.
\end{equation}
The marginals giving qubit observables $\langle I\pm\sigma_i \rangle$
are all non-negative.

The Wigner function is non-negative within the octahedron
$\pm x \pm y \pm z = 1$ embedded in the Bloch sphere (taking into
account both the signs of $\sigma_y$), as illustrated in Fig. 2.
The directions $\hat{n}_j \in \{\pm1,\pm1,\pm1\}$, orthogonal to the faces
of the octahedron, give the maximum negativity to the Wigner function.

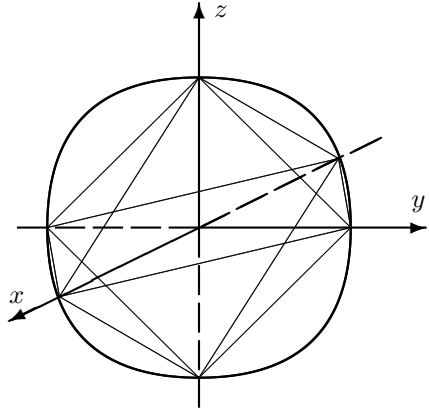
\begin{figure}[t]
\begin{center}
{\setlength{\unitlength}{1cm}
\begin{picture}(8,7)
\thicklines
\put(4,4){\vector(1,0){3}}
\put(4,4){\vector(0,1){3}}
\put(4,4){\vector(-2,-1){2.5}}
\multiput(4,4)(0,-0.5){5}{\line(0,-1){0.4}}
\multiput(4,4)(-0.5,0){5}{\line(-1,0){0.4}}
\multiput(4,4)(0.5,0.25){5}{\line(2,1){0.4}}
\put(1.5,3){\makebox(0,0)[bl]{$x$}}
\put(6.8,4.2){\makebox(0,0)[bl]{$y$}}
\put(4.2,6.8){\makebox(0,0)[bl]{$z$}}

\qbezier(2,4)(2,6)(4,6)
\qbezier(4,6)(6,6)(6,4)
\qbezier(6,4)(6,2)(4,2)
\qbezier(4,2)(2,2)(2,4)

\thinlines
\qbezier(2.16,3.08)(4.08,3.54)(6,4)
\qbezier(2.16,3.08)(2.08,3.54)(2,4)
\qbezier(5.84,4.92)(5.92,4.46)(6,4)
\qbezier(2,4)(3.92,4.46)(5.84,4.92)

\qbezier(4,6)(3,5)(2,4)
\qbezier(4,6)(5,5)(6,4)
\qbezier(4,6)(3.08,4.54)(2.16,3.08)
\qbezier(4,6)(4.92,5.46)(5.84,4.92)

\qbezier(4,2)(3,3)(2,4)
\qbezier(4,2)(5,3)(6,4)
\qbezier(4,2)(3.08,2.54)(2.16,3.08)
\qbezier(4,2)(4.92,3.46)(5.84,4.92)
\end{picture}
}
\end{center}
\vspace{-1.4cm}
\caption{The region of the Bloch sphere corresponding to non-negative
Wigner function for the qubit is the inscribed octahedron. Its vertices
are a unit distance away from the origin along the coordinate axes.}
\end{figure}

Wigner functions for multi-qubit states are easily constructed using
tensor products. For example, the Wigner function for the two-qubit
singlet state becomes:
\begin{equation}
W_{\rm singlet} = {1\over8}\begin{pmatrix}
                  -1 & 1 & 1 & -1\cr 1 & 1 & 1 & 1\cr
                  1 & 1 & 1 & 1\cr -1 & 1 & 1 & -1\end{pmatrix} ~.
\end{equation}
Its negative components are enough to give
$\langle(\vec{\sigma}\cdot\vec{n}_1)(\vec{\sigma}\cdot\vec{n}_2)\rangle
= -\vec{n}_1\cdot\vec{n}_2$, and violate the Bell inequality.

\subsection{Quantum features}

Bell inequalities for experimentally observable correlations are derived
assuming statistical probability distributions for arbitrary local hidden
variables \cite{bell}. Their experimental violation has led to many
discussions about the interpretation of quantum theory. The standard
formulation of quantum theory bypasses them, without introducing any new
variables, when the statistical probability distributions are replaced by
the density matrix. Quantum density matrices bring in complex weights in
general, Wigner functions make the weights real by a specific choice of
representation, but the possibility of the weights being non-probabilistic
(e.g. negative) remains. That is the sense in which Wigner functions are
different from classical phase space distributions.

For a quantum algorithm, Wigner functions can be associated with the initial
product state, the logic gate operations, and the final local measurements.
The outcome probabilities of any quantum evolution can then be expressed as
a phase space probability distribution, which is a product of these Wigner
function factors summed over all evolution time steps $t$ and all quantum
state components $n$. When all the Wigner function factors are non-negative,
the evolution describes a classical stochastic process, which can be
efficiently sampled with an effort polynomial in $n$ and $t$ \cite{eisert}.
This result is robust with respect to sampling errors and bounded
approximations. It generalises the Gottesman-Knill theorem, which states
that all Clifford group quantum operations can be perfectly simulated in
polynomial time on a probabilistic classical computer \cite{gottesman}.

Clifford group operations are those that transform the Pauli group
$\{I,\sigma_x,\sigma_y,\sigma_z\}^{\otimes n}$ within itself, upto phase
factors $\{\pm 1,\pm i\}$. For a single qubit, we can identify them with
the symmetry operations of the octahedron depicted in Fig. 2, which
transform the non-negative Wigner function region to itself:\\
(i) Rotations by angle $\pi$ about an axis through diametrically opposite
vertices flip signs of the transverse components. These are the Pauli
matrix transformations,
$\sigma_j\rightarrow\sigma_i\sigma_j\sigma_i=-\sigma_j$ for $i\ne j$.\\
(ii) Rotations by angles $\pm{\pi\over2}$ about an axis through diametrically
opposite vertices interchange the transverse components (upto a sign). These
are the square-root of Pauli matrix transformations,
$\sqrt{\sigma_i}\sigma_j(\sqrt{\sigma_i})^\dagger$,
and $(\sqrt{\sigma_i})^\dagger\sigma_j\sqrt{\sigma_i}$.\\
(iii) Rotations by angle $\pi$ about an axis through centres of diametrically
opposite edges interchange the edge end-points and flip sign of the third
component, ${1\over\sqrt{2}}(\sigma_i+\sigma_j) \sigma_k
{1\over\sqrt{2}}(\sigma_i+\sigma_j)$.
The Hadamard transformation is of this type.\\
(iv) Rotations by angles $\pm{2\pi\over3}$ about an axis through centres of
diametrically opposite faces cyclically permute the Pauli matrix labels of
the coordinates.\\
(v) The inversion operation flips signs of all three coordinates, and
corresponds to the charge conjugation symmetry of the $SU(2)$ group algebra.\\
Quantum algorithms need something beyond these Clifford group operations to
beat their classical counterparts, which is often achieved by including the
non-Clifford $\root 4 \of \sigma_z$ logic gate.

\section{Quantum Chaos}

In classical dynamics, chaos is characterised as a rapid divergence of
evolution trajectories that are infinitesimally separated to begin with.
Such a divergence makes long-term predictions of a chaotic system
unreliable, when the initial data has limited precision, and the rate
of divergence is specified in terms of the Lyapunov exponents.

A similar description is desirable for quantum systems, to identify whether
they are chaotic or not. Consider two nearby quantum states $\vert\psi\rangle$
and $\vert\psi'\rangle$. Their geodesic separation is specified in terms of
the overlap $\langle\psi\vert\psi'\rangle$, which is invariant under any
unitary evolution
$\vert\psi\rangle \rightarrow U\vert\psi\rangle = e^{-iHt}\vert\psi\rangle$,
and so cannot be used to identify divergent evolution trajectories.
In this context, it has been pointed out that the quantum state should be
treated not as a single point in the Hilbert space, but as an analog of
the phase space distribution of classical systems \cite{qchaos,BinYan}.
The density matrix, with a symplectic structure describable using pairs of
canonically onjugate variables, is the natural setting for such a distribution.
As per the Liouville theorem of canonical classical dynamics, the phase space
density is invariant under evolution. But the quantum state distribution,
spread over an elemental area measured in units of the Planck constant $h$,
can simultaneously stretch in one direction and contract in the conjugate
direction; such a behaviour would characterise chaotic dynamics.

\subsection{Quantum evolution trajectories}

Let the unitary operator $V(t)$ denote the separation between the two
evolution trajectories, i.e. $\vert\psi'(t)\rangle = V(t)\vert\psi(t)\rangle$.
Then,
\begin{equation}
V(t) = e^{-iHt}~V(0)~e^{iHt} ~.
\end{equation}
Parametrising it in terms of a generator direction,
$V(t) = e^{i\epsilon O(t)}$, in the linear regime the evolution obeys:
\begin{equation}
O(t) = e^{-iHt}~O(0)~e^{iHt} ~.
\end{equation}
The corresponding differential evolution equation is:
\begin{equation}
\frac{d}{dt}O(t) = -i[H,O(t)] ~.
\end{equation}
This equation has the same form as the one obeyed by the density matrix
in the Schr\"odinger picture, Eq.(3), and that is not an accident.
(In the Heisenberg picture, where the operators evolve while the states
are held fixed, there is a sign flip in the evolution equations.)
Given a Hamiltonian, the density matrix distribution may evolve to
expand along some directions and contract along some others, and $O(t)$
does the same.

The differential volume element for the symplectic structure of the
density matrix is $dx \wedge dp$ for one-dimensional systems, and
$d(\cos\theta) \wedge d\phi$ for the Bloch sphere. It is quantised in
units of $2\pi$ in the convention where $\hbar=1$. Quantum measurement
limits the information that can be obtained from a symplectic structure,
since only one variable in each conjugate pair of variables can be perfectly
measured. So all the elements of the density matrix are not simultaneously
measurable; only specific projections (e.g. the marginal distributions) are.
The evolution of these projections can be analysed to detect chaos.

For a pure state in an $n$-dimensional Hilbert space, the density matrix is
parametrised by $2n-2$ real parameters. The maximum number of simultaneously
measurable parameters is given by the commuting Cartan subalgebra of $SU(n)$
with $n-1$ generators (which is half of $2n-2$). So just like classical
chaos is defined by the behaviour of the evolution trajectories in the
coordinate space, after projecting out the momenta that form the other half
of the phase space, quantum chaos can be defined by the trajectories of the
Cartan generators. These Cartan generators are fixed by the measurement
operators, and may not commute with the evolution Hamiltonian. (When the
measurement operators commute with the Hamiltonian, there is no evolution
of the observables and no chaos.) In a basis where the Cartan generators
are diagonal, the off-diagonal terms of the Hamiltonian produce transitions
and the density matrix distribution evolves. The transition directions
correspond to raising/lowering operators, and are specified by variables
conjugate to the diagonal ones. The linear response analysis of evolution
along specific generator directions, within the overall evolution that is
unitary, can therefore be used to identify chaos and the Lyapunov exponents.

In this setting, quantum chaos is fully described by the physical
interplay between the measurement basis and the evolution Hamiltonian.
The intrinsic metric of the density matrix space is sufficient for this
purpose, and no other distance measure is needed. Note that entanglement
is also described by the off-diagonal components of the density matrix in
the basis defining the bipartition, and so the same analysis can be extended
to the evolution of entanglement too.

\subsection{Well-known examples}

The inverted harmonic oscillator provides a simple illustration of the
preceding framework. The fundamental commutator is $[x,p]=i$, and we can
choose units such that
\begin{equation}
H = -\frac{1}{2}x^2 + \frac{1}{2}p^2 ~,~~
[H,x] = -ip ~,~~ [H,p] = -ix ~.
\end{equation}
Perturbations in the initial state evolve according to:
\begin{equation}
\frac{dx}{dt} = -p ~,~~ \frac{dp}{dt} = -x ~,
\end{equation}
whose solutions are hyperbolic functions with the Lyapunov exponents $\pm1$
(in contrast to the trigonometric function solutions of the normal harmonic
oscillator with zero Lyapunov exponents).

In the phase space, the evolution matrix for the vector $x \choose p$ is
${\scriptsize \begin{pmatrix} 0&-1\cr -1&0 \end{pmatrix}}$. The absolute
value of its determinant is $1$, which keeps the phase space density constant
as it must.  The Lyapunov exponents are the real parts of the eigenvalues of
this matrix, and the corresponding eigenvectors give the 
expanding/contracting/neutral evolution directions.
The Hamiltonian is a squeezing operator ($H \propto (a^{\dagger2}+a^2)$ in
terms of the creation/annihilation operators), and the evolution exponentially
distorts the phase space distribution. It should be noted in this case that
the exponential separation of initially close trajectories results from the
local maximum in the potential, and is not chaotic or random.

A more relevant example of chaos is provided by the kicked top model,
defined by the Hamiltonian \cite{kickedtop}:
\begin{equation}
H = \frac{\kappa}{2J\tau}J_z^2 + pJ_y\sum_{n=-\infty}^\infty\delta(t-n\tau) ~.
\end{equation}
Here periodic kicks at time interval $\tau$ rotate the state by angle $p$
about the $y$-axis, and $\kappa$ is the chaoticity parameter that twists
the state distribution around the $z$-axis between the kicks. The Floquet
map evolution from kick to kick is given by the unitary operator:
\begin{equation}
U(\tau) = \exp(-i\frac{\kappa}{2J}J_z^2)~\exp(-ipJ_y) ~,
\end{equation}
and the evolution can be represented on the Bloch sphere.

For angular momentum $J$, there are $2J+1$ quantum eigenstates smeared over
the solid angle $4\pi$. The $J\rightarrow\infty$ limit gives the classical
evolution of a point $(\theta,\phi)$ on the Bloch sphere. The transition
between classical and quantum dynamics can be studied by varying $J$.
For $J=\frac{1}{2}$, the Pauli matrix algebra only allows Hamiltonians of
the form $H=b~\hat{n}\cdot\sigma$, which produce only periodic evolution,
i.e. precession of the quantum state around the direction $\hat{n}$.
(There are only two eigenstates, each smeared over a solid angle $2\pi$ of
the Bloch sphere. Such a large smearing eliminates any possibility of chaos.)
For larger $J=1,\frac{3}{2},2,\ldots$, the $J_z^2$ term in the Hamiltonian
contributes to the dynamics, and that is essential for generating chaos.

The connection between classical and quantum state evolution can be
conveniently described using the coherent states $\{\vert\Omega\rangle\}$,
which are obtained by rotating the fully symmetric highest weight state,
$ \vert J,J\rangle = \vert\frac{1}{2},\frac{1}{2}\rangle^{\otimes 2J}$:
\begin{eqnarray}
&& \vert\Omega\rangle = R(\Omega)\vert J,J\rangle ~,~~
R(\Omega) = \exp(i\theta(J_x\sin\phi-J_y\cos\phi)) ~, \cr
&& \langle\Omega\vert\vec{J}\vert\Omega\rangle
= J(\sin\theta\cos\phi~\hat{x},\sin\theta\sin\phi~\hat{y},\cos\theta~\hat{z}) ~.
\end{eqnarray}
In the qubit notation, $\vec{J} = \frac{1}{2}\sum_{i=1}^{2J}\vec{\sigma}^{(i)}$,
\begin{equation}
\vert\Omega\rangle = (\cos\frac{\theta}{2}\vert0\rangle
+ e^{i\phi}\sin\frac{\theta}{2}\vert1\rangle)^{\otimes 2J} ~,
\end{equation}
and the Floquet operator of Eq.(31) can be easily applied to
$\vert\Omega\rangle$ using one- and two-qubit logic gates.

The stereographic projection of the Bloch sphere onto the complex plane
helps in understanding the evolution dynamics.
With $z=\tan\frac{\theta}{2}e^{i\phi}$, the angular momentum eigenstates
are a set of monomials built on the lowest weight state,
$\langle z\vert J,J_z\rangle \propto z^{J+J_z}$, and the group generators are:
\begin{equation}
J_x = \frac{1}{2}(z^2\partial_z - 2Jz -\partial_z)~,~~
J_y = \frac{1}{2i}(z^2\partial_z - 2Jz +\partial_z)~,~~
J_z = z\partial_z - J ~.
\end{equation}
The quantum state evolution is specified by:
\begin{equation}
\frac{d}{dt}z = -i[H,z] = i\frac{\kappa}{2J\tau}(2J-1)z
  - \frac{p}{2}(z^2+1) \sum_{n=-\infty}^\infty\delta(t-n\tau) ~.
\end{equation}
In the Floquet evolution, the term proportional to $\kappa$ gives uniform
rotation, $z \sim \exp(i\frac{\kappa(2J-1)t}{2J\tau})$, as in case of the
harmonic oscillator. The term proportional to $p$ produces phase-dependent
radial jumps, $\Delta(\tanh^{-1}z) \sim -\frac{p}{2}$. Combined together,
they produce a spiral evolution of $z$, and can lead to diverging evolution
trajectories. Chaos requires contribution from both the terms, and it is
absent when $\kappa p(2J-1) = 0$.

\subsection{Tracking chaos using quantum machine learning}

The kicked top model provides a useful strategy to tackle the classification
problem in supervised quantum machine learning. The problem is to efficiently
analyse huge amount of data, collected say by various sensors and detectors,
to make suitable decisions. Often there is no time and space to store the
data, and the interesting features must be extracted quickly while discarding
the rest. Examples cover wide-ranging situations: astronomy, imaging, weather
analysis, collider physics, genetic information and so on.

The typical analysis method is to put together multiple binary classification
steps in a binary tree structure. Each step separates the data with binary
labels into disjoint classes. The classification parameters are first found
using training datapoints with known properties, and then used to determine
the class labels for new datapoints with unknown properties. The capability
of a classifier is enhanced by embedding the datapoints with a nonlinear map
in a larger feature space, $\vec{x}_i \rightarrow \phi(\vec{x}_i)$, and then
performing a linear classification in the feature space; the ideal feature
map would just map the datapoints as $\vec{x}_i\rightarrow\pm1$.

The Hilbert space offers much more versatility in the construction
of feature maps compared to a classical vector space. Using the discrete
logarithm function as a feature map, it has been proved that a quantum
classifier can achieve robust speed-up for a classification problem that is
hard to tackle classically \cite{discretelog}. Furthermore, when the input
classification data originate from a quantum process, they may be directly
fed into a quantum classifier, without intervening measurements that would
project them to classical data. By retaining coherent quantum correlations,
such a procedure can provide an exponential quantum advantage \cite{optbasis}.
In particular, for an $n$-qubit system, expectation values of all $4^n$
elements of its density matrix can be estimated, in the Pauli basis (see
Eq.(37)) upto a constant error, using $O(n)$ copies of the density matrix.
Any classical algorithm must use $2^{\Omega(n)}$ copies of the density matrix
for the same task.

A versatile classifier should be able to construct a variety of structures
covering various patterns of the datapoints, as a function of variational
parameters. The feature map provided by the time evolution of the aperiodic
Heisenberg spin chain Hamiltonian has been found useful for this purpose
(see for example, Ref.\cite{hubregsten}):
\begin{equation}
H = \sum_{\langle i,j\rangle} \alpha_{ij} J_{iz}J_{jz} 
  + \sum_i \beta_i J_{iy} ~.
\end{equation}
In practice, this feature map is implemented as a set of discrete time
evolution Trotter steps, alternating one-qubit rotations with a ring of
two-qubit C-NOT gates as illustrated in Fig. 3. It is easily seen that
the mean-field approximation (i.e. all qubits coupled to each other with
equal strength) to this discrete time evolution is just the kicked top
evolution of Eq.(31). The capability of the latter to go from regular to
chaotic behaviour, as a function of parameter values, hence explains the
success of the former as a versatile feature map.

\begin{figure}[t]
{
\begin{center}
\begin{picture}(320,90)
\put(0,78){$\vert 0\rangle$}
\put(0,58){$\vert 0\rangle$}
\put(0,38){$\vert 0\rangle$}
\put(0,18){$\vert 0\rangle$}
\put(0,-2){$\vert 0\rangle$}

\put(20,80){\line(1,0){15}}
\put(20,60){\line(1,0){15}}
\put(20,40){\line(1,0){15}}
\put(20,20){\line(1,0){15}}
\put(20,0){\line(1,0){15}}

\put(35,75){\framebox(10,10){$H$}}
\put(35,55){\framebox(10,10){$H$}}
\put(35,35){\framebox(10,10){$H$}}
\put(35,15){\framebox(10,10){$H$}}
\put(35,-5){\framebox(10,10){$H$}}

\put(45,80){\line(1,0){15}}
\put(45,60){\line(1,0){15}}
\put(45,40){\line(1,0){15}}
\put(45,20){\line(1,0){15}}
\put(45,0){\line(1,0){15}}

\put(60,75){\framebox(30,10){$R(x_1)$}}
\put(60,55){\framebox(30,10){$R(x_2)$}}
\put(60,35){\framebox(30,10){$R(x_3)$}}
\put(60,15){\framebox(30,10){$R(x_4)$}}
\put(60,-5){\framebox(30,10){$R(x_5)$}}

\put(90,80){\line(1,0){15}}
\put(90,60){\line(1,0){15}}
\put(90,40){\line(1,0){15}}
\put(90,20){\line(1,0){15}}
\put(90,0){\line(1,0){15}}

\put(105,75){\framebox(30,10){$R_y(\theta_1)$}}
\put(105,55){\framebox(30,10){$R_y(\theta_2)$}}
\put(105,35){\framebox(30,10){$R_y(\theta_3)$}}
\put(105,15){\framebox(30,10){$R_y(\theta_4)$}}
\put(105,-5){\framebox(30,10){$R_y(\theta_5)$}}

\put(135,80){\line(1,0){105}}
\put(135,60){\line(1,0){15}}
\put(135,40){\line(1,0){60}}
\put(135,20){\line(1,0){15}}
\put(135,0){\line(1,0){60}}

\put(240,75){\framebox(35,10){$R_z(\theta_{10})$}}
\put(150,55){\framebox(30,10){$R_z(\theta_6)$}}
\put(195,35){\framebox(30,10){$R_z(\theta_7)$}}
\put(150,15){\framebox(30,10){$R_z(\theta_8)$}}
\put(195,-5){\framebox(30,10){$R_z(\theta_9)$}}

\put(275,80){\line(1,0){10}}
\put(180,60){\line(1,0){105}}
\put(225,40){\line(1,0){60}}
\put(180,20){\line(1,0){105}}
\put(225,0){\line(1,0){60}}

\put(165,80){\circle*{4}}
\put(165,80){\line(0,-1){15}}
\put(210,60){\circle*{4}}
\put(210,60){\line(0,-1){15}}
\put(165,40){\circle*{4}}
\put(165,40){\line(0,-1){15}}
\put(210,20){\circle*{4}}
\put(210,20){\line(0,-1){15}}
\put(255,0){\circle*{4}}
\put(255,0){\line(0,1){75}}

\put(100,-10){\dashbox{5}(180,100){}}
\put(290,82){\line(1,0){2}}
\put(290,-2){\line(1,0){2}}
\put(292,-2){\line(0,1){84}}
\put(300,38){$\vert\phi(x_i)\rangle$}
\end{picture}
\end{center}
}
\vspace{5mm}
\caption{An illustrative digital quantum logic circuit for constructing
the feature map $x_i\rightarrow\phi(x_i)$. The boxed part of the logic
circuit is iterated several times to yield the Trotter decomposed evolution
of the aperiodic Heisenberg spin chain Hamiltonian, with the variational
parameters $\theta_j$. The initial part of the logic circuit inputs the
datapoint parameters in the uniform superposition state.}
\end{figure}
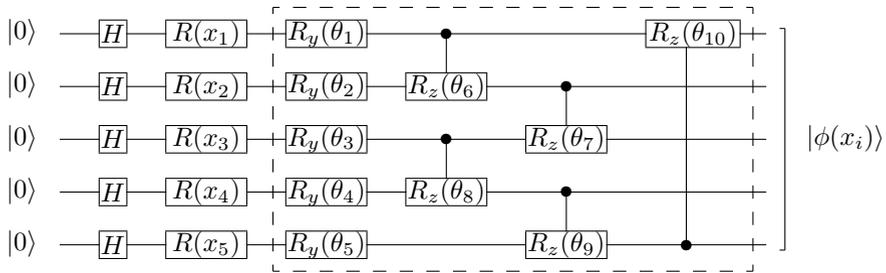

We can now demonstrate the power of quantum machine learning by converting
the kicked top evolution to a binary classification problem, and showing
that it can be efficiently solved in both regular and chaotic regimes.
Let the kicked top evolve for time $n\tau$, starting from the initial
coherent state $\vert\Omega\rangle$. Given the final state, the binary
classification task is to predict whether the initial state was in the
northern or the southern hemisphere of the Bloch sphere. The solution
requires finding an (approximate) inverse evolution map.

\begin{figure}[t]
\includegraphics[width=6cm]{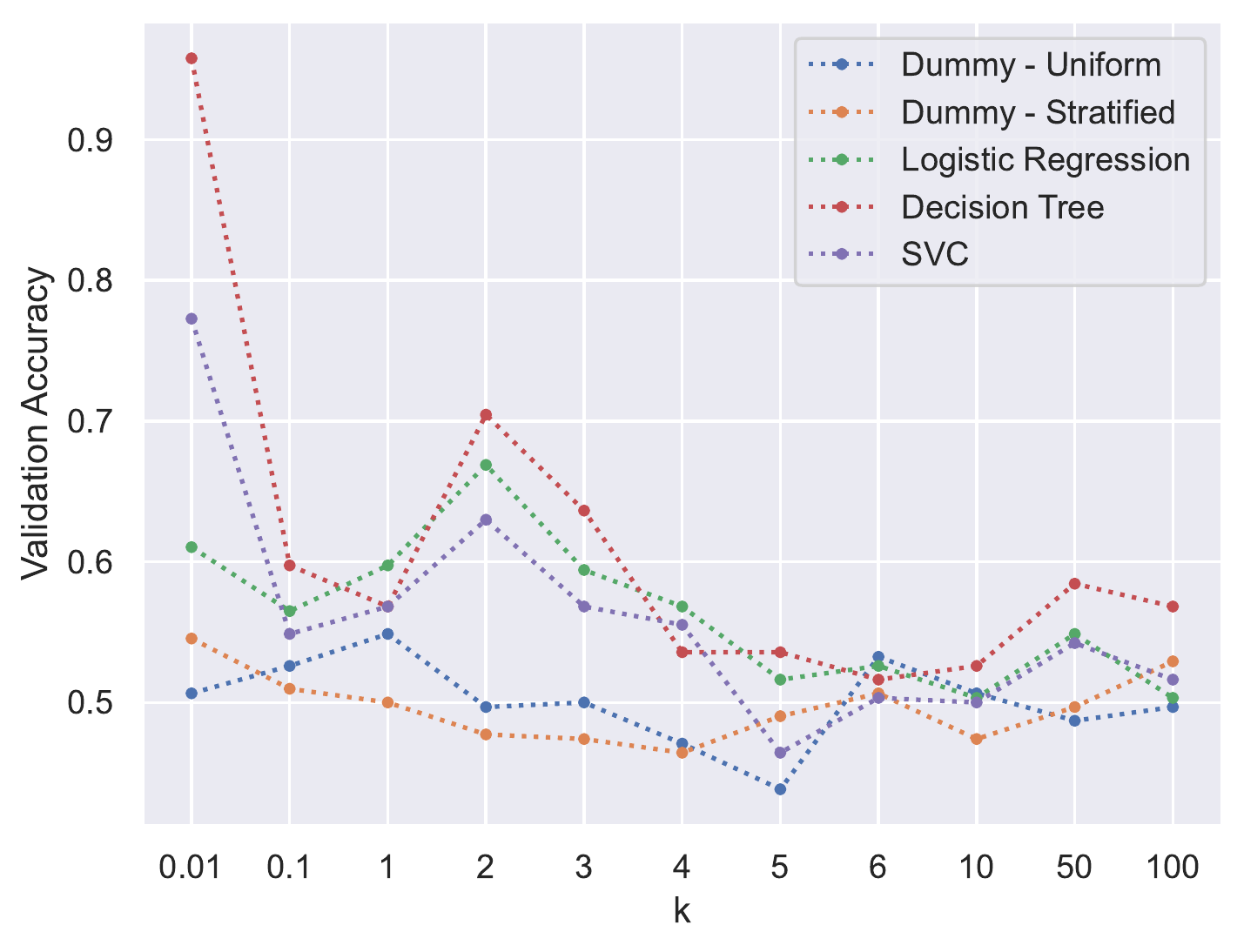}
\includegraphics[width=6cm]{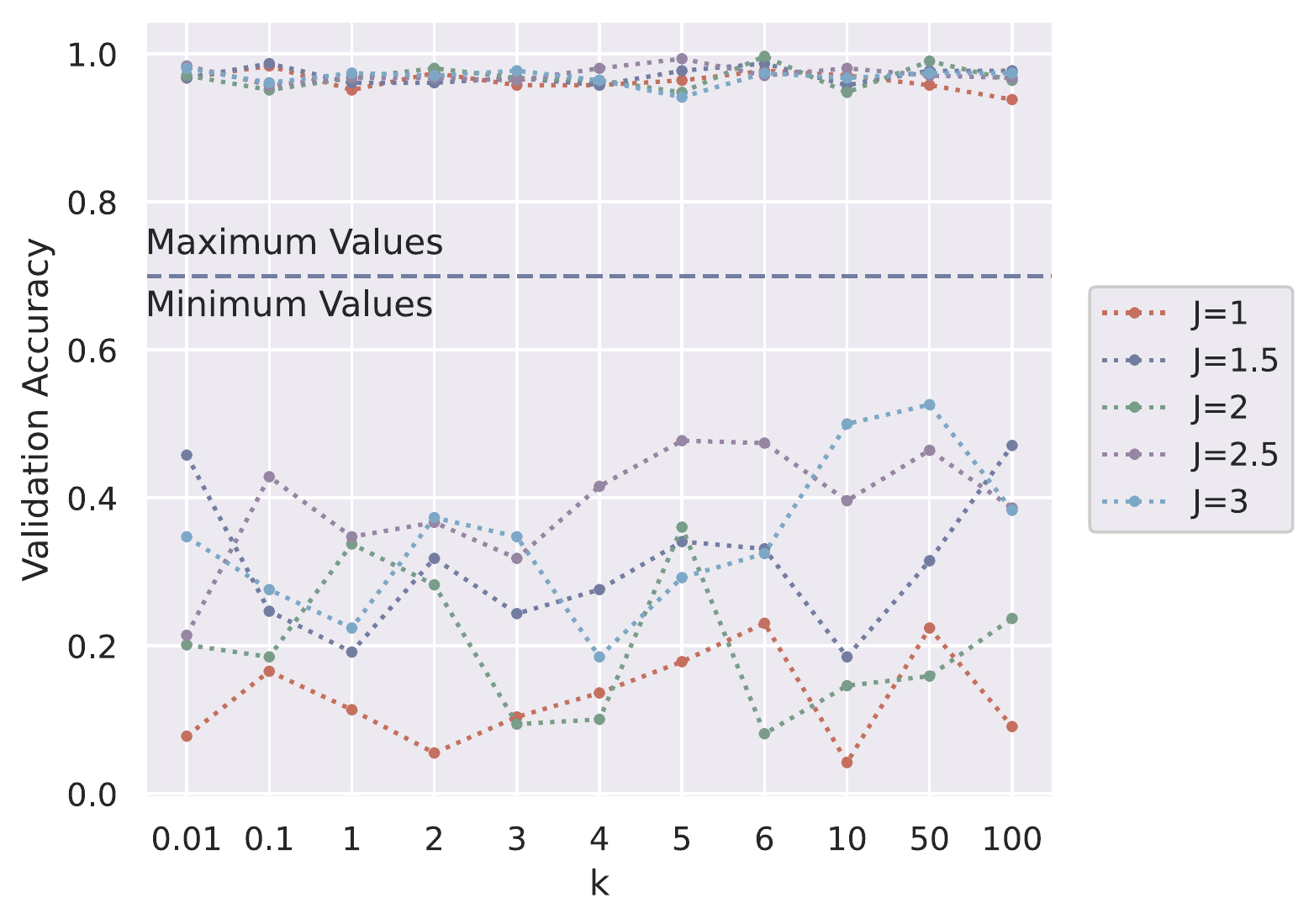}

\vspace{-2mm}
\hspace{3.1cm}(a)\hspace{5.1cm}(b)\hfill
\vspace{2mm}
\caption{Comparison of classical and quantum classifier performances for
the evolution of a kicked top: (a) The success rates (validation accuracy)
of several common classical machine learning methods as a function of the
chaoticity parameter $\kappa$. (b) The success rates of the aperiodic
Heisenberg spin chain form quantum classifier, as a function of $\kappa$
for $J$ varying from $1$ to $3$. The minimum and maximum values indicate
the results before and after tuning the variational parameters $\theta_j$
using the training datapoints.}
\end{figure}

The capabilities of backtracking the kicked top evolution were compared
between classical and quantum machine learning methods using numerical
simulations \cite{ankit}. For ease of simulation, the rotation parameter
$p$ was set to $\frac{\pi}{2}$, while varying the chaoticity parameter
$\kappa$. The Bloch sphere was uniformly discretised as $32\times32$
datapoints on the $(\theta,\phi)$ grid, and the number of time evolution
steps $n$ ranged from $1$ to $1000$. $30\%$ of the datapoints were randomly
selected as the training datapoints, and the rest were used to check the
success rate of the trained classification method. In the case of classical
evolution, the initial $\vert\Omega\rangle$ corresponds to a point on the
Bloch sphere. The evolution changes from being regular at small $\kappa$
to chaotic for large $\kappa$, with the transition to chaos occurring at
$\kappa=4$. As illustrated in Fig. 4a, the success rate of various classical
machine learning methods decreases with increasing $\kappa$. The accuracy for
correctly predicting the starting hemisphere of the evolution is non-trivial
for small $\kappa$, but becomes essentially the random guess value $0.5$
once $\kappa$ crosses over to the chaotic regime.

In the case of quantum evolution, the Floquet operator of Eq.(31) evolves
the initial coherent state $\vert\Omega\rangle$ of $2J$ qubits for $n$ time
steps, and the resultant state is directly fed into the classification
logic circuit (i.e. the boxed part of Fig. 3) without any measurement.
After executing the classification logic circuit for $l$ iterations, the
class prediction is chosen as the sign of the expectation value of the first
qubit. The probabilistic nature of quantum measurement means that the whole
algorithm has to be run several times, say $m$, to determine the expectation
value to a reasonable accuracy. In the numerical simulations, both $l$ and
$m$ were kept finite, around $10$. The training datapoints were used to tune
the variational parameters of the classifier, $\theta_j$, so as to maximise
the success rate of class prediction. The results are shown in Fig. 4b for
$J$ varying from $1$ to $3$; while training the variational parameters is
essential for optimising the success rate, once that is done, the class
prediction is highly successful for all values of $\kappa$ and $J$. The
cross-over of $\kappa$ from regular to chaotic regime, or a large difference
between $n$ and $l$, have no discernible effect. This is a striking result.
The successful backtracking of the kicked top evolution with a finite depth
classifier is due to the fact that the initial coherent quantum state is
smeared over a solid angle $\frac{4\pi}{2J+1}$ on the Bloch sphere, instead
of being a point. The lesson is that the smearing of quantum states in the
phase space suppresses chaos as well as makes it possible to backtrack it.

\section{Noisy Quantum Processor Simulation}

Quantum systems are highly sensitive to disturbances from the environment;
even necessary controls and observations perturb them. The available, and
upcoming, quantum devices are noisy, and techniques to bring down the
undesired errors are being intensively pursued. This era of noisy
intermediate scale quantum systems has been labeled NISQ \cite{NISQ}.
It is also necessary to come up with error-resilient system designs,
as well as techniques that validate and verify the results. Such NISQ
systems roughly span devices with 10-100 qubits, 10-1000 logic operations,
limited interactions between qubits, and with no error correction since
the fault-tolerance threshold is orders of magnitudes away. They would
likely be used as special purpose platforms, with limited capabilities.

Software simulators are being developed to help in investigations of
noisy quantum processors \cite{qcsimulators}. They are programs running
on classical parallel computer platforms, which are designed to mimic
noisy quantum processors, and can model and benchmark 10-50 qubit systems.
A quantum computation may suffer from many sources of error: imprecise
initial state preparation, imperfect logic gate execution, disturbances
to the data in memory, and error-prone measurements. (It is safe to assume
that the program instructions, which are classical, are essentially
error-free.) A realistic quantum simulator needs to include all of them
with appropriate probability distributions. Additional features that can
be included are restrictions on possible logic operations and connectivity
between the components, which would imitate what may be the structure of a
real quantum processor. With such improvisations, the simulation results
would look close to what a noisy quantum processor would deliver, and one
can test how well various algorithms work with imperfect quantum components.
More importantly, one can vary the imperfections and the connectivity in
the software simulator to figure out what design for the noisy quantum
processor would produce the best results.

Quantum simulators serve an important educational purpose as well. They
are portable, and can be easily distributed over existing computational
facilities worldwide. They provide a platform to students to acquire
the skills of {\em programming} as well as {\em designing} quantum
processors, which is of vital importance for developing future expertise
in the field.

\subsection{Generic implementation}

The standard formulation of quantum states as vectors in a Hilbert space
evolving by unitary transformations is appropriate for describing the pure
states of a closed quantum system, but is insufficient for describing the
mixed states that result from interactions of an open system with its
environment. The evolution of generic mixed states is described using
the density matrix formulation, the CPTP map of Eq.(12), where various
environmental disturbances are modeled by suitable choices of the Kraus
operators $\{M_\mu\}$. It is an ensemble description of the quantum system,
and so is inherently probabilistic, in contrast to the deterministic state
vector description that can describe individual experimental system evolution.
Nonetheless, it allows determination of the expectation value of any physical
observable, which is the average result over many experimental realisations.

In going from a description based on $\vert\psi\rangle$ to the one based
on $\rho$, the number of degrees of freedom gets squared. This property
is fully consistent with the Schmidt decomposition, which implies that any
correlation between the system and the environment can be specified by
modeling the environment using a set of degrees of freedom as large as
that for the system. The squaring of the degrees of freedom is the price
to be paid for the flexibility to include all possible environmental
effects on the quantum system, and it slows down the classical simulation
of an open quantum system.

Consider computational problems whose algorithms have already been converted
to discrete quantum logic circuits acting on a set of qubits. Also assume
that all logic gate instructions can be executed with a fixed clock step.
In this framework, the computational complexity of the program is specified
by the number of qubits and the total number of clock steps. Since the
quantum state deteriorates with time due to environmental disturbances,
the total execution time is reduced by identifying non-overlapping logic
operations at every clock step and then implementing them in parallel.

The density matrix of an $n$-qubit quantum register can be expressed in the
orthogonal Pauli basis, utilising the tensor product structure of the Hilbert
space, as
\begin{equation}
\rho = \sum_{i_1,i_2,\ldots,i_n} a_{i_1 i_2\ldots i_n}
(\sigma_{i_1}\otimes\sigma_{i_2}\otimes\ldots\otimes\sigma_{i_n}) ~.
\end{equation}
Here $i_1,\ldots,i_n\in\{0,1,2,3\}$, $\sigma_0\equiv I$, and
$a_{i_1 \ldots i_n}$ are $4^n$ real coefficients encoded as an array.
The normalisation Tr$(\rho)=1$ implies $a_{0\ldots0}=2^{-n}$.
The constraint Tr$(\rho^2)\le1$, which follows from $\rho^2\preceq\rho$,
implies $\sum_{i_1,\ldots,i_n} a_{i_1 \ldots i_n}^2 \le 2^{-n}$.
The orthogonality of the Pauli basis makes it easy to describe various
transformations of the density matrix as simple changes of the coefficients.
(When the density matrix is expressed as a $2^n\times2^n$ complex Hermitian
matrix, the number of independent components remain the same, but the matrix
elements do not belong to an orthogonal set, and their transformations do not
have the same type of compact description.) It has also been observed, due
to the fact that all the Pauli basis elements mutually either commute or
anticommute, that the Pauli basis is highly efficient for actual quantum
hardware measurements \cite{optbasis}.

When the operator to be measured is expressed in the same Pauli basis,
\begin{equation}
O = 2^{-n}\sum_{i_1,i_2,\ldots,i_n} o_{i_1 i_2\ldots i_n}
(\sigma_{i_1}\otimes\sigma_{i_2}\otimes\ldots\otimes\sigma_{i_n}) ~,
\end{equation}
its expectation value is just the dot product,
\begin{equation}
{\rm Tr}(\rho O)
= 2^{-n}\sum_{i_1,i_2,\ldots,i_n} a_{i_1 i_2\ldots i_n}o_{i_1 i_2\ldots i_n} ~.
\end{equation}
Also, treating the coefficients $a_{i_1\ldots i_n}$ as vector space
coordinates, the density matrices can be discriminated in terms of the
Euclidean distance between them; the Hilbert-Schmidt distance is just the
$L_2$ distance in the $4^n$-dimensional space:
\begin{equation}
{\rm Tr}((\rho_1-\rho_2)^2) = 2^{-n} \sum_{i_1,i_2,\ldots,i_n}
(a_{i_1 i_2\ldots i_n}-b_{i_1 i_2\ldots i_n})^2
= 2^{-n} \vert\vert\vec{a}-\vec{b}\vert\vert^2_2 ~.
\end{equation}

The reduced density matrix with the degrees of freedom of the $k^{\rm th}$ qubit
summed over, ${\rm Tr}_k(\rho)$, is specified by the $4^{n-1}$ coefficients
$2 a_{i_1\ldots i_{k-1} 0 i_{k+1}\ldots i_n}$, since only the terms containing
$\sigma_0$ provide a non-zero partial trace. Upon a projective measurement,
the quantum state components orthogonal to the direction of measurement 
vanish. So when the $k^{\rm th}$ qubit is measured along direction $\hat{n}$,
the coefficients $a_{\ldots i_k\ldots}$ are set to zero for $i_k\perp\hat{n}$,
while those for $i_k\vert\vert\hat{n}$ and $i_k=0$ remain unchanged.

\subsection{The QSim simulator}

In the framework of the preceding subsection, we constructed a simulator
for noisy quantum logic circuits \cite{Qsim}. It is an open-source software
library written in Python \cite{aakash}, which is added as a new backend to
IBM's Qiskit platform \cite{qiskit}. It has been made freely available as
a national educational resource \cite{qctoolkit}.

We consider problems where all operations---logic gates, errors and
measurements---are local, i.e. act on only a few qubits. Indeed, the Qiskit
transpiler decomposes more complicated operations into a sequence of one-qubit
and two-qubit operations. The tensor product structure of such operations is
a non-trivial operator on the addressed qubits and the identity operator on
the rest of the qubits. Since the expression for the quantum register has the
same tensor product structure, the operations change the Pauli matrix factors
corresponding to only the addressed qubits (e.g. $\sigma_{i_k}$),
and the coefficients change only for the associated subscripts (e.g.
$a_{\ldots i_k\ldots})$. Such operations are efficiently implemented in the
software using linear algebra vector instructions (there is no complex number
algebra in our code), while explicitly evaluating Eq.(12).

The manipulations of logic circuit instructions and operations are carried
out at the classical level; even when a quantum hardware is available,
they would be executed by a classical compiler. So we assume that they are
error-free. We incorporate possible errors in initialisation, logic gates,
measurement and memory using simple models:\\
$\bullet$
We allow a fully factorised thermal state as one of the initial state options:
\begin{equation}
\rho_{\rm th} = \begin{pmatrix}
                p & 0 \cr 0 & 1-p \cr
                \end{pmatrix}^{\otimes n} ,
\end{equation}
where the parameter $p$ is provided by the user.\\
$\bullet$
For single qubit rotations around fixed axes, we assume that errors arise
from inaccuracies in their rotation angles. Let $\alpha$ be the inaccuracy
in the angle, with the mean
$\langle\langle\alpha\rangle\rangle=\overline{\alpha}$ and the fluctuations
symmetric about $\overline{\alpha}$. Then the replacement
$\theta\rightarrow\theta+\alpha$ in the rotation operator $R_n(\theta)$
modifies the density matrix transformation according to the substitutions:
\begin{equation}
\cos\theta \rightarrow r\cos(\theta+\overline{\alpha}) ~,~~
\sin\theta \rightarrow r\sin(\theta+\overline{\alpha}) ~,
\end{equation}
where $\overline{\alpha}$ and
$r = \langle\langle\cos(\alpha-\overline{\alpha})\rangle\rangle$
are the parameters provided by the user.\\
$\bullet$
To model the error in the C-NOT gate, we assume that C-NOT is implemented
as a transition selective pulse that exchanges amplitudes of the two target
qubit levels when the control qubit state is $\vert1\rangle$, for example as
in NMR quantum information processors. Then the error is in the duration of
the transition selective pulse, and alters only the second half of the unitary
operator,
$U_{cx} = \vert0\rangle\langle0\vert \otimes I
+ \vert1\rangle\langle1\vert \otimes \sigma_1$.
It is included in the same manner as the error in single qubit rotation
angle (i.e. as a disturbance to the rotation operator $\sigma_1$).
The corresponding two parameters, analogous to $\overline{\alpha}$ and
$r$, are provided by the user.\\
$\bullet$
We model a single qubit projective measurement error as depolarisation,
which is equivalent to a bit-flip error in a binary measurement.
Then when the $k^{\rm th}$ qubit is measured along direction $\hat{n}$,
the coefficients $a_{\ldots i_k\ldots}$ in the post-measurement state are
set to zero for $i_k\perp\hat{n}$, reduced by a multiplicative factor $d_1$
(provided by the user) for $i_k\vert\vert\hat{n}$, and left unaffected for
$i_k=0$. Also, the probabilities of the two outcomes become
$\frac{1}{2}(1\pm 2^n d_1\hat{n}\cdot\vec{c})$, where
$\{c_0,\vec{c}\} = a_{0\ldots i_k\ldots0}$. In case of a measurement of
a multi-qubit Pauli operator string, the above procedure is applied to
every qubit whose measurement operator has $i_k\ne0$.\\
$\bullet$
In case of a Bell-basis measurement of qubits $k$ and $l$, the
post-measurement coefficients with $i_k\ne i_l$ are set to zero, those
with $i_k=i_l\in\{1,2,3\}$ are reduced by a multiplicative factor $d_2$
provided by the user, and those with $i_k=i_l=0$ are left the same.
Also, the probabilities of the four outcomes are obtained by reducing
the $i_k=i_l\in\{1,2,3\}$ contributions by the factor $d_2$.\\
$\bullet$
We assume that the memory errors are small during a clock step, and
implement them by modifying the density matrix at the end of every clock
step, in the spirit of the Trotter expansion. With the $\sigma_3$ basis
as the computational basis, the decoherence effect suppresses the
off-diagonal coefficients with $i_k\in\{1,2\}$ for every qubit by a
multiplicative factor $f$. It can be represented by the Kraus operators:
\begin{equation}
M_0 = \sqrt{\frac{1+f}{2}}~I ~,~~ M_1 = \sqrt{\frac{1-f}{2}}~\sigma_3 ~.
\end{equation}
In terms of the clock step size $\Delta t$ and the decoherence time $T_2$,
the parameter $f=\exp(-\Delta t/T_2)$, and it is provided by the user.\\
$\bullet$
We consider the decay of the quantum state towards the thermal state,
Eq.(41). This evolution is represented by the Kraus operators:
\begin{eqnarray}
M_0 &=& \sqrt{p}\begin{pmatrix}
                1 & 0 \cr 0 & \sqrt{g} \cr\end{pmatrix} ,~
M_1  =  \sqrt{p}\begin{pmatrix}
                0 & \sqrt{1-g} \cr 0 & 0 \cr\end{pmatrix} ,\\
M_2 &=& \sqrt{1-p}\begin{pmatrix}
                  \sqrt{g} & 0 \cr 0 & 1 \cr\end{pmatrix} ,~
M_3  =  \sqrt{1-p}\begin{pmatrix}
                  0 & 0 \cr \sqrt{1-g} & 0 \cr\end{pmatrix} .\nonumber
\end{eqnarray}
Its effect on every qubit is to suppress the off-diagonal coefficients
with $i_k\in\{1,2\}$ by $\sqrt{g}$, and change the diagonal coefficients
according to:
\begin{equation}
a_{\ldots3\ldots} \rightarrow
g~a_{\ldots3\ldots} + (2p-1)(1-g) a_{\ldots0\ldots} ~.
\end{equation}
In terms of the clock step $\Delta t$ and the relaxation time $T_1$,
the parameter $g=\exp(-\Delta t/T_1)$, and it is provided by the user.
(Our Kraus representation automatically ensures the physical constraint
$T_2\le2T_1$). We note that the decoherence and decay superoperators
commute with each other, and we execute the combined operation at the
end of every clock step.

We expect the memory errors to cause maximum damage to the quantum signal,
because they act on all the qubits all the time, while other operational
errors are confined to particular qubits at specific instances. Our tests
for simple algorithms confirm this expectation, and so we consider it
imperative to reduce the total execution time of a quantum program as much
as possible. Towards this end, we rearrange the complete list of quantum
logic circuit instructions produced by the Qiskit transpiler in a set
of partitions, such that all operations in a single partition can be
executed as parallel threads during a single clock step. In the process,
successive single qubit rotations are merged wherever feasible, a stack of
sequential operations is constructed for every qubit, and non-overlapping
qubit operations are collected in a single partition wherever possible.
This procedure puts logic gate operations and measurement operations in
separate partitions, since single qubit measurement may affect the whole
quantum register in case of entangled quantum states. Also, the clock step
is assumed to be longer than the time required to execute each operation
in the corresponding partition.

In a quantum computer, elementary logic operations would be directly executed
on quantum hardware, and the computational complexity of the algorithm is
specified in terms of the number of qubits and the number of elementary logic
operations required. In a classical simulator, the elementary logic operations
are executed using linear algebra, and so its computational complexity is
specified in terms of the number of linear algebra operations (permutations,
additions and multiplications) required to execute various elementary logic
operations. For our simulator, these computational resource requirements are
easily enumerated:\\
$\bullet$ Memory: Storage of an $n$-qubit register needs $4^n$ real
variables.\\
$\bullet$ Logic gates: One- and two-qubit operations are block-diagonal
matrices with fixed block sizes. Their execution requires $O(4^n)$ linear
algebra operations.\\
$\bullet$ Measurement: The probability of a measurement outcome for an
observable is just a value look up in the density matrix simulator. For an
$n$-qubit register, it is an $O(n)$ effort.\\
$\bullet$ Environmental noise: The one-qubit Kraus operators parametrising
the noise are block-diagonal matrices with fixed block sizes. They are
implemented using $O(4^n)$ linear algebra operations.

Overall, in going from a quantum state vector simulator to a density matrix
one, the computational resource requirements increase from $O(2^n)$ to
$O(4^n)$. It implies that given certain computational resources, a density
matrix simulator can simulate half the number of qubits compared to a state
vector one. On the other hand, a density matrix simulator produces the
complete output probability distribution in one run, while a state vector
simulator requires multiple runs (labeled shots) of the program for the same
purpose. Our fully portable Qsim simulates quantum logic circuits with $10$
qubits and $100$ operations in a few minutes on a laptop.

The main achievement of our simulator is the ability to simulate noisy
quantum systems, using simple error models. By varying the noise parameter
values, we can estimate how accurately we need to control various errors in
quantum hardware in order to get meaningful results. Simulation of simple
algorithms shows that the errors produced by decoherence and decay cause
the largest deterioration of the results, with decay dominating over
decoherence \cite{Qsim}.

\section{Outlook}

Quantum theory was invented because classical theory could not explain
certain observations at all. Quantum technology therefore can be advantageous
when these phenomena are at the core of the problems to be tackled; they
include superposition, entanglement, squeezing and tunnelling of quantum
states. The quantum density matrix provides a complete description of quantum
states, generalising the classical concept of probability distribution by
adding extra degrees of freedom (as off-diagonal matrix elements). These
extra degrees of freedom cover genuinely quantum phenomena that often appear
to be non-intuitive. The unitary transformations available in quantum logic
are more powerful than their subgroups of permutation operations available in
classical reversible logic. For computational problems with classical initial
and final states, they can provide shortcuts through the Hilbert space,
leading to more efficient algorithms. Such strategies are vigorously being
pursued, even in absence of a systematic procedure.

In addition, two noteworthy research directions at present involving the
quantum density matrix are:
(i) The technique of classical shadows \cite{shadows}, which quantifies how
much can be learned about a quantum system using efficient classical methods
(and hence what are the quantum features that would be hard to extract).
(ii) Quantum machine learning \cite{qml}, which uses efficient quantum
feature maps that are hard to construct classically for speeding up
classification and data analysis problems.

\bmhead{Acknowledgments}

This work was supported in part by the project ``Centre for Excellence in
Quantum Technology", funded by the Ministry of Electronics and Information
Technology, Government of India.

\section*{Declarations}

\begin{itemize}
\item The corresponding author states that there is no conflict of interest.
\end{itemize}

\bibliography{sn-bibliography}


%

\end{document}